\documentclass[aps,prd,floatfix,nofootinbib]{revtex4} 

\usepackage{graphics}
\usepackage{graphicx}

\usepackage{latexsym}
\usepackage[cp866]{inputenc}
\usepackage[english]{babel}

\input epsf

\begin{document}

\title{\boldmath $\eta(1295)\to3\pi$ decays}
\author{N.N. Achasov \footnote{achasov@math.nsc.ru}
and G.N. Shestakov \footnote{shestako@math.nsc.ru}
}\affiliation{Laboratory of  Theoretical Physics, S.L. Sobolev
Institute for Mathematics, 630090, Novosibirsk, Russia}


\begin{abstract}
The radially excited pseudoscalar state $\eta(1295)$ is still poorly
understood [as well as its $SU(3)$ partners $\pi(1300)$ and $K(1460
)$] and we want to attract attention experimenters to its
comprehensive study. As for the main three-body decay $\eta(1295)\to
\eta\pi\pi$, here the main interest is the measurements of the
shapes of the $\eta\pi $ and $\pi\pi$ mass spectra in which the
contributions from the $a_0(980)$ resonance and the large $\pi\pi$
$S$ wave ($\sigma$) should manifest themselves. To describe these
mass spectra, we propose to use a simple isobar model with a single
fitting parameter characterizing the relative intensity of the $a_0
(980)$ and $\sigma$ state production. Our main goal is to discuss
the dynamics of the isospin-breaking decays $\eta(1295)\to\pi^+
\pi^-\pi^0$ and $\eta(1295)\to3\pi^0$, whose experimental studies
could continue the impressive story of isospin violations in the
decays of light isoscalar mesons $\omega\to2\pi$, $\eta\to3 \pi$,
$\eta'\to3\pi$, $\eta(1405)\to3\pi$, and $f_1(1285)\to3\pi$. We
estimate the widths of the isospin-breaking decays $\eta(1295)\to
3\pi$ produced via the mixing of the $\pi^0-\eta$ and $a^0_0(980)
-f_0(980)$ states and also due to the $\pi^0(1300)-\eta(1295)$
mixing. Processes that have the potential for detecting $\eta(1295)
\to3\pi$ decays are discussed.
\end{abstract}

\maketitle

\section{Introduction}

Mechanisms of the isospin-breaking decays of light mesons are highly
diverse. For instance, the decay of $\omega\to\pi^+\pi^-$
\cite{PDG2020} occurs mainly due to the $\rho^0-\omega$ mixing
\cite{GFQ69,GSR69}. The seed mechanism of the $\eta\to3\pi$
\cite{Lo07,Am08,Pr18} and $\eta'\to3\pi$ \cite{Na09,Ab17} decays is
the $\pi^0-\eta$ mixing, the manifestation of which is significantly
enhanced due to the strong interaction of pions in the final state.
This fact was gradually elucidated as a result of great efforts over
more than fifty years to explain the data on the decays $\eta\to
3\pi$, see Refs. \cite{RT81,OW70,GTW79,GL85,Ho02,BB03,BMN06,Col18,
Gan20,KW96,AL96,Da15,Da17,Ol20,Ol20a,FKK21} for details. In these
works, the complicated technique is presented for taking into
account pair interactions in three-pion final states based on the
unitarized chiral perturbation theory and solutions of dispersion
equations for the $\pi\pi$ $S$ and $P$ waves. Decays $a^0_0(980)
\to\pi^+\pi^-$ and $f_0(980)\to \eta\pi^0$ discovered by the BESIII
Collaboration \cite{Ab11,Ab18} are due to the $a^0_0(980)-f_0(980)$
mixing \cite{ADS79,AS04a,AS04b,AS19}. The significant isospin
breaking in the $\eta(1405)\to f_0(980)\pi^0\to3\pi$ decays is
explained mainly by the triangle logarithmic singularity, which is
present in the transition amplitude $\eta(1405)\to(K^*\bar K+\bar
K^*K)\to(K^+K^-+ K^0\bar K^0)\pi^0\to f_0(980)\pi^0\to3\pi$
\cite{Ab12,WLZZ12,ALQWZ12, WWZZ13,AKS15,AS18,DZ19}. Experimental
results on the search for $a^0_0(980)-f_0(980)$ mixing in the
$\eta(1405)\to f_0(980)\pi^0 \to3\pi$ decays \cite{Ab12} and also in
the $f_1(1285)\to f_0(980)\pi^0\to3\pi$ decays, which were
discovered by the VES \cite{Do08,Do11} and BESIII \cite{Ab15}
Collaborations, suggest a broader perspective on the isotopic
symmetry breaking effects due to the $K^+$ and $K^0$ mass
difference. It has become clear that not only the $a^0_0(980)-f_0
(980)$ mixing but also any mechanism producing $K\bar K$ pairs with
a definite isospin in an $S$-wave gives rise to such effects
\cite{AKS16,AS19}, thus suggesting a new tool for studying the
nature and production mechanisms of light scalars. As for the
$\eta(1295)\to3\pi$ decays, a hint at their existence was obtained
by the BESIII Collaboration in the study of the processes $e^+e^-\to
J/\psi\to\gamma3\pi$ \cite{Ab12}. However, it remains unclear which
of the two mesons $\eta(1295)$ or $f_1(1285)$ (or both together)
lead to a slight excess over the smooth background in the three-pion
mass spectra around 1290 MeV \cite{Ab12}.

In the present paper, we discuss the mechanisms that can lead to the
isospin-breaking decays $\eta(1295)\to\pi^0\pi^+\pi^-$ and $\eta(129
5)\to3\pi^0$, and also the processes having the potential for
detecting such decays. The paper is organized as follows. In Sec. II
we construct a simple isobar model for the main decay of the $\eta(1
295)$ into $\eta\pi\pi$. The model takes into account the amplitudes
of subprocesses $\eta(1295)\to a_0(980)\pi\to\eta\pi\pi$ and $\eta(1
295)\to\eta\sigma\to\eta(\pi\pi)_{{\scriptsize S}}$ the coherent sum
of which defines the shapes of the $\eta\pi$ and $\pi\pi$ mass
spectra or the corresponding $\eta(1295)\to\eta\pi\pi$ Dalitz plots
(here $\sigma$ is a symbolic notation for the virtual $S$-wave
hadronic system with isospin $I=0$ decaying into $\pi\pi$). Fitting
the data on the $\eta\pi$ and $\pi\pi $ mass spectra within this
model using a single parameter characterizing the relative intensity
of the $a_0(980)$ and $\sigma$ state production will allow easily to
elucidate whether it is necessary to use for their description more
complex theoretical models (similar, for example, to those discussed
in Refs. \cite{Gan20,FKK21} for $\eta'\to\eta\pi\pi$ decays). Note
that unlike the decays $\eta\to3\pi$, $\eta'\to3 \pi$, and $\eta'\to
\eta\pi\pi$ any theoretical predictions relative to the overall
normalization of the decay amplitude $\eta(1295)\to\eta \pi\pi$ is
absent. That is, with a modern state of the theory, this
normalization can be determined only from the experimental data on
the decay width of $\eta(1295)\to\eta\pi\pi$. In Sec. III we use the
above model for $\eta(1295)\to\eta\pi\pi$ to estimate the widths of
the direct decays $\eta(1295)\to3\pi$ caused by the $\pi^0-\eta$ and
$a^0_0(980)-f_0(980)$ mixing. These sources of the isospin breaking
lead to very different shapes of the two-pion mass spectra. Due to
the $\pi^0-\eta$ mixing, the $\pi^0\pi^+$, $\pi^+\pi^-$, and $\pi^0
\pi^0$ mass spectra turn out to be wide and rather smooth. The
$a^0_0(980)-f_0(980)$ mixing amplitude is large between the $K^+K^-$
and $K^0\bar K^0$ thresholds \cite{ADS79} and as a result leads to
the narrow (about 10 MeV wide) resonancelike structures in $\pi^+
\pi^-$ and $\pi^0\pi^0$ mass spectra in the 1 GeV energy region.
Another source of $\eta(1295)\to3\pi$ decays is the mixing of the
$\eta(1295)$ and $\pi^0(1300)$ resonances. The presence of this
mechanism, which violates isospin, fundamentally distinguishes the
case under consideration from the cases of the decays $\eta\to3\pi
$, $\eta'\to3\pi$, $\eta(1405)\to3\pi$, and $f_1(1285)\to3\pi$. In
Sec. IV we discuss the effect of the $\pi^0(1300)-\eta(1295)$ mixing
and present estimates for the $\eta(1295)\to3\pi$ decay widths
caused by the considered isospin-breaking mechanisms. Note that when
estimating the widths of the direct decays $\eta (1295)\to3\pi$
caused by $\pi^0-\eta$ and $a^0_0(980)-f_0(980)$ mixing, we did not
take into account the rescattering effects of pions in the final
state. However, such approach utilizing a minimum number of free
parameters presents a quite reasonable guide for the primary
treatment of future data. Currently, data on the $\eta(1295)\to3\pi$
decays is completely absent, and the applicability of the chiral
perturbation theory to the analysis of $\eta(1295)\to3\pi$ is not
obvious. By the way, the transition $\eta(1295)\to\pi^0(1300)\to
3\pi$ can be considered as a peculiar kind of final state
interaction. In Sec. V the processes that can be used to search for
$\eta(1295)\to3\pi$ decays are discussed. The results of our
analysis are briefly summarized in Sec. VI.

\section{The \boldmath{$\eta(1295)\to\eta\pi^+\pi^-$} decay}

First of all, we note that the $\eta(1295)$ meson, like its probable
$SU(3)$ partners $\pi(1300)$ and $K(1460)$ \cite{PDG2020}, has not
been sufficiently studied yet, despite a large number of experiments
performed \cite{PDG2020,AM20,St79,An86,Bi88,Fu91,Au92,Al97,Te98,
Ma03,Ad01,Au08}. Progress in the investigation of the $\eta(1295)$
would be highly desirable, especially since its searches in
$\gamma\gamma$ collisions \cite{Acc01}, in central production
\cite{Bar97,Bar98} and in a number of experiments on the radiative
$J/\psi$ decays \cite{Bai99,Bai00,Bai04} did not give the expected
results. Recently, the properties of excited pseudoscalar states
have been discussed in Refs. \cite{Au08,PG17, PdA17,Aai18,CZ21,
FL21}.

The state $\eta(1295)$ was first discovered as a result of a
partial-wave analysis of the $\eta\pi^+\pi^-$ system in the reaction
$\pi^-p\to\eta\pi^+\pi^-n$ at 8.45 GeV \cite{St79} and then
confirmed in other experiments on the reactions $\pi^-p\to\eta\pi^+
\pi^-n$ \cite{An86,Fu91,Te98,Ma03}, $\pi^-p\to\eta\pi^0\pi^0n$
\cite{Al97}, $\pi^-p\to K^+K^0_S\pi^-n$ \cite{Bi88}, $\pi^-p\to K^+
K^-\pi^0n$ \cite{Ad01}, $J/\psi\to\gamma\eta\pi^+\pi^-$ \cite{Au92},
and $B^+\to K^+\eta\pi\pi$ \cite{Au08} (see also \cite{PDG2020,
AM20}). In almost all experiments, the separation of signals from
$\eta(1295 )$ and $f_1(1285)$ states, having common decay modes, was
carried out. The results of the partial-wave analyzes indicate that
the $\eta(1295)$ decays predominantly via quasi-two-body
intermediate states: $\eta(1295)\to a_0(980)\pi\to\eta\pi\pi$ and
$\eta(1295)\to\eta\sigma\to\eta(\pi\pi)_{{\scriptsize S}}$. The
relation between the $\eta(1295)\to a_0(980)\pi$ and $\eta
(1295)\to\eta\sigma$ modes are not well defined \cite{PDG2020,AM20,
St79,An86,Bi88,Fu91,Au92,Al97,Te98,Ma03,Ad01,Au08}. But there are no
special indications of the dominance of any one of them. According
to the Particle Data Group (PDG) \cite{PDG2020}, the mass and total
width of the $\eta(1295) $ are $1294\pm4$ MeV and $55\pm5$ MeV,
respectively .

To estimate the probabilities of the isospin-breaking decays
$\eta(1295)\to3\pi$, it is necessary to have a model for the main
decay $\eta(1295)\to\eta\pi^+\pi^-$. The decay width of $\eta(1295)
\to\eta\pi^0\pi^0$ is related to the $\eta(1295)\to\eta\pi^+\pi^-$
one by the isotopic relation $2\Gamma_{\hat{\eta}\to\eta\pi^0
\pi^0}=\Gamma_{\hat{\eta}\to\eta \pi^+\pi^+}$ [here and hereinafter
$\hat{\eta}$ is a short notation for $\eta(1295)$]. We use a simple
isobar model (see, for example, Refs. \cite{Ab17,Am94,Ab14} and
references therein) and write the amplitude of the
$\eta(1295)\to\eta\pi^+\pi^-$ decay  as follows:
\begin{eqnarray}\label{Eq1} F_{\hat{\eta}\to\eta\pi^+\pi^-}(s,t,u)
=T_{a^+_0}(s)+T_{a^-_0}(t)+T^{\hat{\eta}}_{\sigma}(u),\end{eqnarray}
where $s$, $t$, and $u$ are the $\eta\pi^+$, $\eta\pi^-$, and $\pi^+
\pi^-$ invariant mass squared, respectively, and the amplitudes $T$
have the form
\begin{eqnarray}\label{Eq2}T_{a^+_0}(s)+T_{a^-_0}(t)=\frac{g_{\hat{
\eta}a_0\pi}g_{a_0\eta\pi}}{16\pi}\left(\frac{1}{D_{a^+_0}(s)}+
\frac{1}{D_{a^-_0}(t)}\right), \qquad T^{\hat{\eta}}_{\sigma}(u)
=C_{\hat{\eta}}\,T^0_0(u)=C_{\hat{\eta}}\,\frac{\eta^0_0(u)e^{
\delta^0_0(u)}-1}{2i\rho_{\pi^+\pi^-}(u)}.
\end{eqnarray}
That is, the amplitudes $T_{a^+_0}(s)$ and $T_{a^-_0}(t)$ are
saturated by the contributions of intermediate $a^+_0(980)$ and
$a^-_0(980)$ states that manifest themselves in the $\eta(1295)\to
\eta\pi^+\pi^-$ decay in the form of peaks in the $\eta\pi^+$ and
$\eta\pi^-$ mass spectra, respectively; here $D_{a^\pm_0}$ is the
inverse propagator of the $a^\pm_0(980)$ resonance in which the
finite width corrections are taken into account (see its explicit
form, for example, in Refs. \cite{AKS16, AS19,AKS21}). For the
coupling constants of the $a_0(980)$ with pairs of light
pseudoscalar mesons, we use the relations valid in the four-quark
model \cite{Ja77,AI89}:
\begin{eqnarray}\label{Eq3}
g_{a_0\eta\pi}=\bar{g}\cos(\theta_i-\theta_p),\ \
g_{a_0\eta'\pi}=-\bar{g}\sin(\theta_i-\theta_p),\ \ g_{a^+_0K^+
K^0}=g_{a^-_0K^0 K^-}=\bar{g},\end{eqnarray} where $\bar{g}$ is the
overall coupling constant, $\theta_i=35. 3^\circ$ is the so-called
``ideal'' mixing angle and $\theta_p=-11.3^\circ$ is the mixing
angle in the nonet of the light pseudoscalar mesons \cite{PDG2020}.
The corresponding decay widths have the standard form $\sqrt{s}
\Gamma_{a_0\to ab}(s)=g^2_{a_0ab}\rho_{ab}(s)/(16\pi)$, where
$\rho_{ab}(s)=[s^2-2s(m^2_a+m^2_b)+(m^2_a-m^2_b)^2 ]^{1/2}/s$. For
further estimates, we set $m_{a_0}=0.985$ GeV and $g^2_{a_0
\eta\pi}/(16\pi)=0.2$ GeV$^2$ \cite{AKS16,AS19, AKS21}. The
amplitude $T^{\hat{\eta}}_{\sigma}(u)$, describing the interaction
in the $\pi\pi$ channel, is taken according to Eq. (\ref{Eq2})
proportional to the $S$-wave $\pi\pi$ scattering amplitude with
isospin $I=0$; $\eta^0_0(u)$ and $\delta^0_0(u)$ are its
inelasticity (equal to 1 for $u<4m^2_{K^+}$) and phase,
respectively. We take the amplitude $T^0_0(u)$ from Ref. \cite{AK06}
at the values of the parameters indicated in Table I for fitting
variant I. In this work, the excellent simultaneous descriptions of
the phase shifts, inelasticity, and mass distributions in the
reactions $\pi\pi\to\pi\pi$, $\pi\pi\to K\bar K$, and $\phi\to
\pi^0\pi^0 \gamma$ was obtained.

\begin{figure} 
\begin{center}\includegraphics[width=14cm]{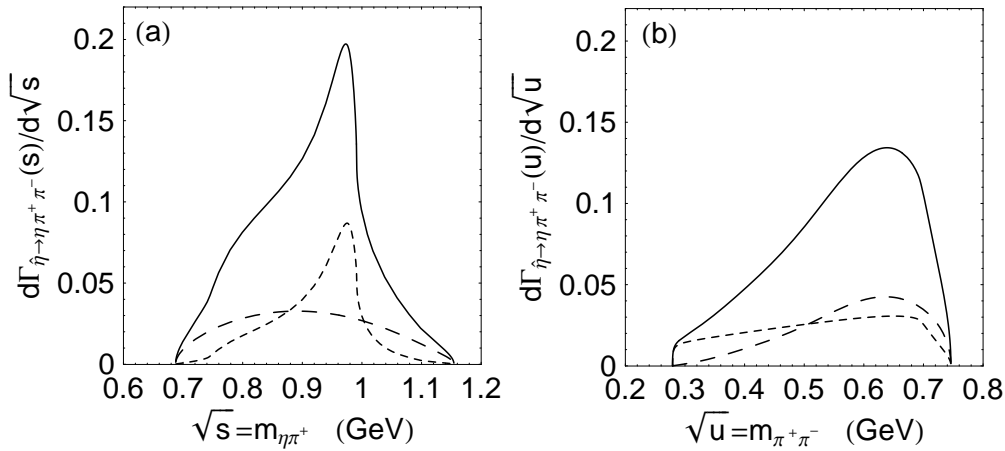}
\caption{\label{Fig1} The solid curves show the mass spectra (a)
$\eta\pi^+$ and (b) $\pi^+\pi^-$ in the decay $\eta(1295)\to\eta
\pi^+ \pi^-$. Contributions from the $a_0(980)\pi$ and $\eta\sigma$
intermediate states are shown by short and long dashed curves,
respectively.} \end{center}\end{figure}

The variables $s$, $t$, and $u$ are related by the relation
$\Sigma\equiv s+t+u=M^2+m^2_\eta+2m^2_{\pi^+}$, where $M$ is the
invariant mass of the initial (virtual) state $\eta(1295)$. To
simplify the notations, we do not indicate $M$ among the arguments
on which the amplitudes of the considered decays depend. Choosing
$s\equiv m^2_{\eta\pi^+}$ and $u\equiv m^2_{\pi^+\pi^-}$ as
independent variables and taking into account the adopted
normalizations in Eqs. (\ref{Eq1}) and (\ref{Eq2}), we write the
total decay width of $\eta(1295)\to\eta\pi^+\pi^-$ in the form
\begin{eqnarray}\label{Eq4} \Gamma_{\hat{\eta}
\to\eta\pi^+\pi^-}=\frac{1}{\pi M^3}\int\left|F_{\hat{\eta}\to
\eta\pi^+\pi^-}(s,t,u)\right|^2dsdu\,.\end{eqnarray} Integration
limits for three-body decays are given in Ref. \cite{MT21}. The
constants $g_{\hat{\eta}a_0\pi}$ and $C_{\hat{\eta}}$ introduced in
Eq. (\ref{Eq2}) can be estimated by assuming the condition of
equality of the contributions from the amplitudes $T_{a^+_0}(s)+
T_{a^-_0}(t)$ and $T^{\hat{\eta}}_{\sigma}(u)$ into the $\eta(1295)
\to\eta\pi^+\pi^-$ decay width (see the discussion of experimental
data in the second paragraph of this section) and normalizing the
contribution of the module squared of their coherent sum [see Eqs.
(\ref{Eq1}) and (\ref{Eq4}] to the value $\Gamma_{\hat{\eta}\to\eta
\pi^+\pi^-}\approx(2/3)\Gamma^{\mbox{\scriptsize{tot}}}_{\hat{\eta}
}\approx36$ MeV. For the nominal mass of the $\eta(1295)$ meson
\cite{PDG2020}, we obtain $g_{\hat{\eta}a_0 \pi}\approx1.26$ GeV and
$C_{\hat{\eta}}\approx0.626$. The mass spectra of the $\eta\pi^+$
and $\pi^+\pi^-$ pairs,
\begin{eqnarray}\label{Eq5}
\frac{d\Gamma_{\hat{\eta}\to\eta\pi^+\pi^-}(s)}{d\sqrt{s}}=
\frac{2\sqrt{s}}{\pi M^3}\int\left|F_{\hat{\eta}\to
\eta\pi^+\pi^-}(s,t,u)\right|^2du\ \ \ \mbox{and}\ \ \
\frac{d\Gamma_{\hat{\eta}\to\eta\pi^+\pi^-}(u)}{d\sqrt{u}}=
\frac{2\sqrt{u}}{\pi M^3}\int\left|F_{\hat{\eta}\to
\eta\pi^+\pi^-}(s,t,u)\right|^2ds\,,\end{eqnarray} corresponding to
the above values of the parameters are shown in Fig. \ref{Fig1}. The
peak due to the $a^+_0(980)$ resonance is clearly visible in the
$\eta\pi^+$ mass spectrum. The mass spectrum of $\pi^+\pi^-$ is
naturally smoother. By varying the constants $g_{\hat{\eta}a_0\pi}$
and $C_{\hat{\eta}}$, one can obtain various shapes for these mass
spectra and use the specified parametrization for fitting.
Unfortunately, for the $\eta(1295)\to\eta\pi^+\pi^-$ decay there is
still no data cleared of significant foreign admixtures. We are now
in position to move on to estimating probabilities of the decays
$\eta(1295)\to3\pi$.

\section{Direct decays \boldmath{$\eta(1295)\to\pi^0\pi^+\pi^-$ and $\eta(1295)\to3\pi^0$}}

\begin{figure} [!ht] 
\begin{center}\includegraphics[width=14cm]{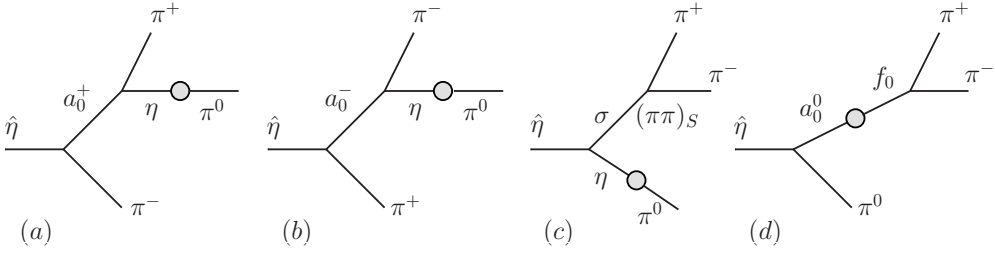}
\caption{\label{Fig2a} Direct decays $\eta (1295)\to
\pi^0\pi^+\pi^-$.}
\end{center}\end{figure}

The diagrams responsible for the direct decays $\eta(1295)\to\pi^0
\pi^+\pi^-$ are shown in Fig. \ref{Fig2a}. Diagrams (a), (b), and
(c) are due to the $\pi^0-\eta$ mixing \cite{GTW79}, and diagram (d)
is due to the $a^0_0(980)-f_0(980)$ mixing \cite{ADS79}. The
corresponding amplitude can be written in the form
\begin{eqnarray}\label{Eq6}
F^{\mbox{\scriptsize{dir}}}_{\hat{\eta}\to\pi^0\pi^+\pi^-}(s,t,u)\equiv
F^{\mbox{\scriptsize{dir}}}_c(s,t,u)
=\frac{\Pi_{\pi^0\eta}}{m^2_\eta-m^2_{\pi^0}}\left(T_{a^+_0}(s)+
T_{a^-_0}(t)+T^{\hat{\eta}}_{\sigma}(u)\right)+\frac{g_{\hat{
\eta}a_0\pi}g_{f_0\pi^+\pi^-}}{16\pi}\,e^{i\delta_B(u)}G_{a^0_0f_0}(u),
\end{eqnarray}
where $\Pi_{\pi^0\eta}$ is the mass squared of the $\pi^0-\eta$
transition, $\delta_B(u)$ is a smooth and large phase (of about
$90^\circ$ for $\sqrt{u}=m_{ \pi^+\pi^-}\approx1$ GeV) of the
elastic background accompanying the $f_0(980)$ resonance in the
$S$-wave reaction $\pi\pi\to\pi\pi$ in the channel with isospin
$I=0$ \cite{ADS79,AS04a,AS04b}, and $G_{a^0_0f_0}(u)$ is the
propagator of the $a^0_0(980)-f_0(980)$ transition
\cite{ADS79,AS19,AKS16};
\begin{eqnarray}\label{Eq7} G_{a^0_0f_0}(u)=
\frac{\Pi_{a^0_0f_0}(u)}{D_{a^0_0}(u)D_{f_0}(u)-\Pi^2_{
a^0_0f_0}(u)}\,,\end{eqnarray}
\begin{eqnarray}\label{Eq8}
\Pi_{a^0_0f_0}(u)=\frac{g_{a^0_0K^+K^-}g_{f_0K^+K^-}}{16\pi}
\left[i[\rho_{K^+K^-}(u)-\rho_{K^0\bar K^0}(u)]-\frac{\rho_{K^+K^-}
(u)}{\pi}\,\ln\frac{1+\rho_{K^+K^-}(u)}{1-\rho_{K^+K^-}(u)}\right.
\nonumber \\ \left.+ \frac{\rho_{K^0\bar K^0}(u)}{\pi}\,\ln\frac{1+
\rho_{K^0\bar K^0}(u)}{1-\rho_{K^0\bar K^0}(u)}\right], &&
\end{eqnarray}
where  $\rho_{K\bar K}(u)=\sqrt{1-4m^2_K /u}$ for $\sqrt{u}\geq2m_K
$; if $\sqrt{u}\leq2m_K$, then $\rho_{K\bar K}(u)$ should be
replaced to $i|\rho_{K\bar K}(u)|$. In Eq. (\ref{Eq7}) $D_{f_0}(u)$
is the inverse propagator of the $f_0(980)$ resonance coupled with
$\pi\pi$, $K\bar K$, and $\eta\eta$ channels (see, for example, Ref.
\cite{AKS16}). For further estimates, we set $m_{f_0}= 0.985$ GeV,
$(3/2)g^2_{f_0 \pi^+\pi^-}/(16\pi)=0.098$ GeV$^2$, $2 g^2_{f_0
K^+K^-}/(16\pi)=0.4$ GeV$^2$, and $g^2_{f_0\eta\eta} =g^2_{ f_0
K^+K^-}$ \cite{AKS16}. Notice that the phase of the amplitude of the
$a^0_0(980)-f_0(980)$ mixing, $\Pi_{a^0_0f_0}(u)$, in the region
between  $K^+K^-$ and $K^0\bar K^0$ thresholds changes by about
$90^\circ$ \cite{AKS16,AS04a,AS04b}. This fact is crucial for the
observation of the $a^0_0(980)-f_0(980)$ mixing effect in
polarization experiments \cite{AS04a,AS04b}. According to the
analysis presented in Ref. \cite{Fe00}, we use to estimate
$\Pi_{\pi^0\eta}$ the value equal to $-0.004 $ GeV$^2$, see also
Refs. \cite{Io01,Ac19}.

\begin{figure} [!ht] 
\begin{center}\includegraphics[width=14cm]{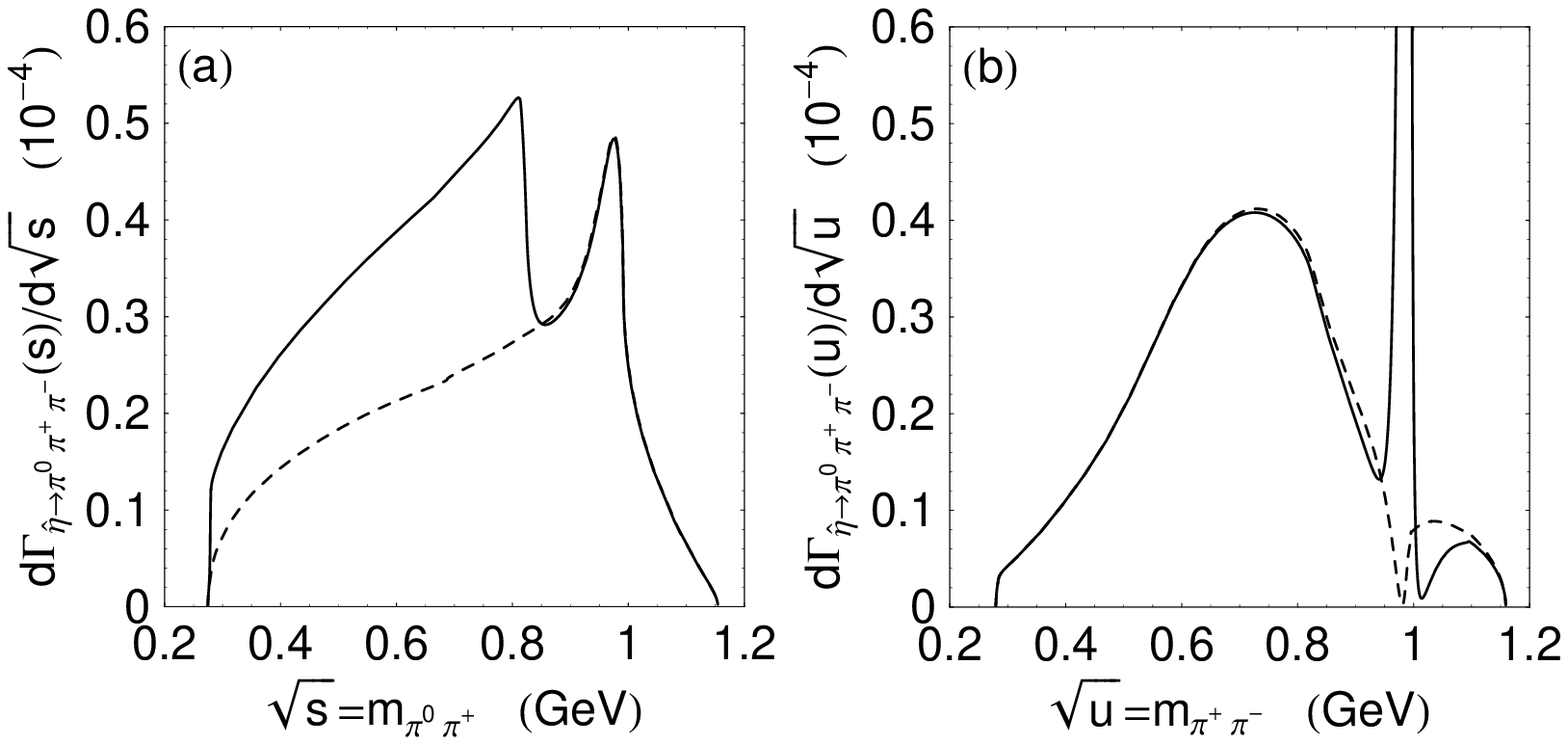}
\caption{\label{Fig2} The solid curves show (a) $\pi^0\pi^+$ and (b)
$\pi^+\pi^-$ mass spectra in the $\eta(1295)\to\pi^0\pi^+ \pi^-$
decay; due to the $a^0_0(980)-f_0(980)$ mixing mechanism the $\pi^+
\pi^-$ mass spectrum reaches $\approx11$ at its maximum. The areas
under the solid curves correspond to $\Gamma^{
\mbox{\scriptsize{dir}}}_{\hat{\eta}\to\pi^0\pi^+\pi^-}=0.027$ MeV.
The dashed curves show the contributions due to $\pi^0-\eta $
mixing. }\end{center}\end{figure}

\begin{figure} [!ht] 
\begin{center}\includegraphics[width=7cm]{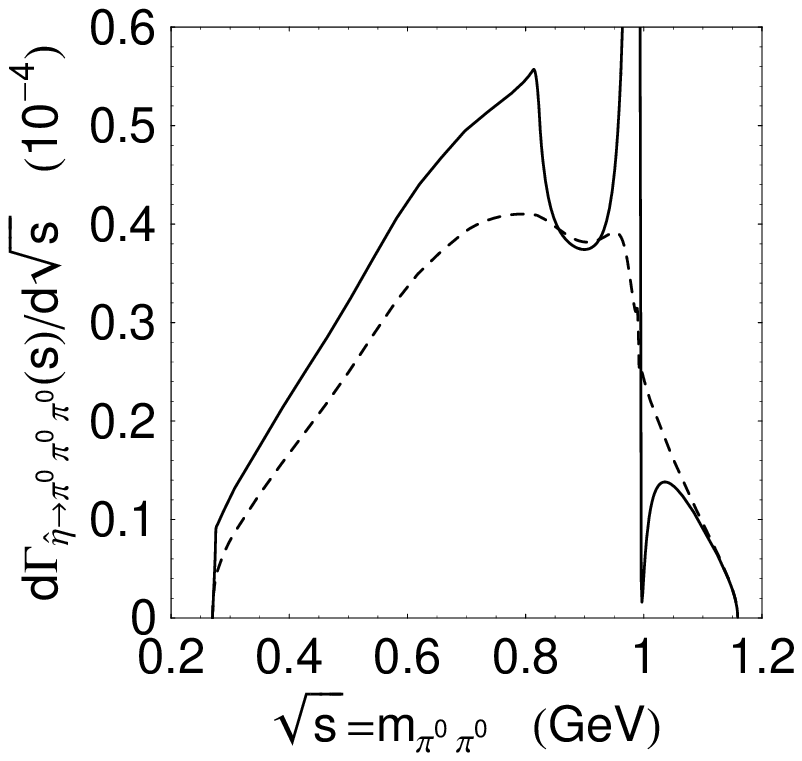}
\caption{\label{Fig3} The solid curve shows $\pi^0\pi^0$ mass
spectrum in the $\eta(1295)\to3\pi^0$ decay; due to the
$a^0_0(980)-f_0(980)$ mixing mechanism the $\pi^0\pi^0$ mass
spectrum reaches $\approx3.4$ at its maximum. The area under the
solid curve corresponds to $\Gamma^{\mbox{\scriptsize{dir}}
}_{\hat{\eta}\to3 \pi^0}=0.031$ MeV. The dashed curve shows the
contribution due to $\pi^0-\eta $ mixing. }\end{center}\end{figure}

\begin{figure} 
\begin{center}\includegraphics[width=14cm]{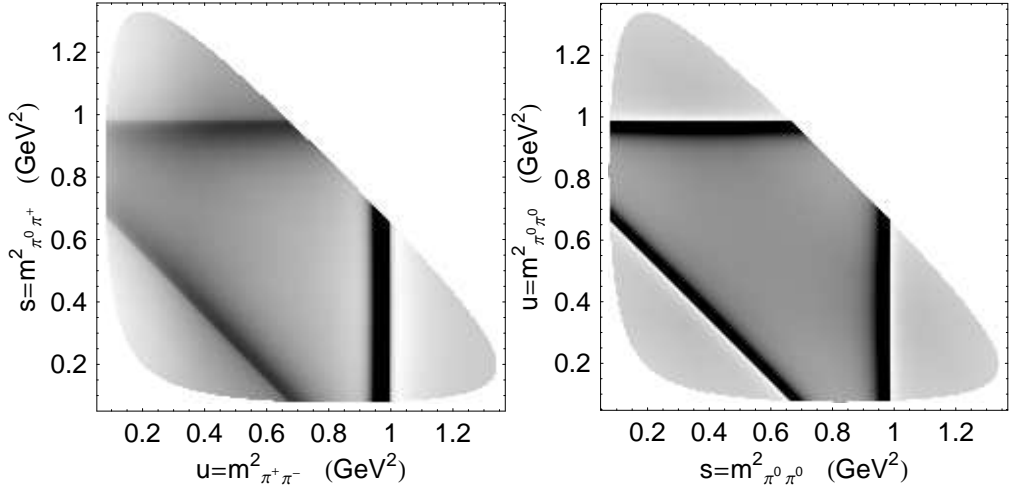}
\caption{\label{Fig4} The Dalitz plots for distributions (a)
$|F^{\mbox{\scriptsize{dir}}}_c(s,t,u)|^2/(\pi M^3)$ and (b)
$|F^{\mbox{\scriptsize{dir}}}_n(s,t,u)|^2/(6\pi M^3)$. The $a^0_0
(980)-f_0(980)$ mixing mechanism is responsible for the areas of the
strongest blackening in these plots. }\end{center}\end{figure}

The decay amplitude $\eta(1295)\to3\pi^0$, taking into account the
identity of the $\pi^0$ mesons, can be written as \cite{Col18}
\begin{eqnarray}\label{Eq9}
F^{\mbox{\scriptsize{dir}}}_{\hat{\eta}\to3\pi^0}(s,t,u)\equiv
F^{\mbox{\scriptsize{dir}}}_n(s,t,u)=F^{\mbox{\scriptsize{dir}}}_c(s,t,u)+
F^{\mbox{\scriptsize{dir}}}_c(u,s,t,)+F^{\mbox{\scriptsize{dir}}}_c(t,u,s).
\end{eqnarray} The full decay widths
$\Gamma^{\mbox{\scriptsize{dir}}}_{\eta(1295)\to\pi^0\pi^+\pi^-}$
and $\Gamma^{\mbox{\scriptsize{dir}}}_{\eta (1295)\to3\pi^0}$ are
\begin{eqnarray}\label{Eq10} \Gamma^{\mbox{\scriptsize{dir}}}_{\hat{\eta}
\to\pi^0\pi^+\pi^-}=\frac{1}{\pi M^3}\int\left|
F^{\mbox{\scriptsize{dir}}}_c(s,t,u)\right|^2dsdu\ \ \ \mbox{and}\ \
\ \Gamma^{\mbox{\scriptsize{dir}}}_{\hat{\eta} \to3\pi^0}=\frac{1}{
6\pi M^3}\int\left| F^{\mbox{\scriptsize{dir}}}_n(s,t,u)\right|^2
dsdu\,.\end{eqnarray} For the above values of the parameters, we get
\begin{eqnarray}\label{Eq11}
\Gamma^{\mbox{\scriptsize{dir}}}_{\hat{\eta}\to\pi^0\pi^+\pi^-}\approx0.027\
\mbox{MeV}\ \ \ \mbox{and}\ \ \
\Gamma^{\mbox{\scriptsize{dir}}}_{\hat{\eta}\to3\pi^0}\approx0.031 \
\mbox{MeV}. \end{eqnarray} As can be seen from Figs. \ref{Fig2} and
\ref{Fig3}, both mechanisms breaking isospin, $\pi^0-\eta$ mixing
and $a^0_0(980)-f_0(980)$ mixing, make significant contributions to
$\Gamma_{\hat{\eta}\to3\pi}$. The contribution from the $a^0_0(980)-
f_0(980)$ mixing to the $\pi^+\pi^-$ and $\pi^0\pi^0$ mass spectra
is concentrated in a narrow region near the $K\bar K$ thresholds,
see Figs. \ref{Fig2}(b) and \ref{Fig3}. At the maximum, this
contribution reaches $\approx11$ in the $\pi^+ \pi^-$ mass spectrum
[Fig. \ref{Fig2}(b)] and $\approx3.4$ in the $\pi^0\pi^0$ one (Fig.
\ref{Fig3}). For completeness, Fig. \ref{Fig4} shows the Dalitz
plots for distributions $|F^{\mbox{\scriptsize{dir}}}_c(s,t,u)|^2
/(\pi M^3)$ and $|F^{\mbox{\scriptsize{dir}}}_n(s,t, u)|^2/(6\pi
M^3)$. The $a^0_0(980)-f_0(980)$ mixing mechanism is responsible for
the areas of the strongest blackening in these plots.

\section{\boldmath{$\pi^0(1300)-\eta(1295)$ mixing}}

Suppose that for the states $\eta(1295)$ and $\pi^0(1300)$ that are
close in mass, their mixing occurs at the same level as the $\rho^0-
\omega$ mixing \cite{PDG2020,GFQ69,GL82,Ur95,Ak02}. Thus we set
Re$\Pi_{\hat{\pi}^0\hat{\eta}}\approx\mbox{Re}\Pi_{\rho^0\omega}
\approx\Pi_{\rho^0\omega}\approx0.0034$ GeV$^2$, where
$\Pi_{\hat{\pi}^0\hat{\eta}}$ and $\Pi_{\rho^0\omega}$ are the
square of the messes of the $\pi^0(1300)-\eta(1295)$ and $\rho^0-
\omega$ transitions, respectively [here and hereinafter $\hat{\pi}$
is a short notation for $\pi(1300)$]. Hence, the following rough
estimate can be obtained for the width of the $\eta(1295)\to3\pi$
decay caused by the $\pi^0(1300)-\eta(1295)$ mixing:
\begin{eqnarray}\label{Eq12} \Gamma^{\mbox{\scriptsize{mix}}}_{\hat{\eta}
\to\hat{\pi}^0\to3\pi}\approx\left|\frac{\mbox{Re}\Pi_{\hat{\pi}^0
\hat{\eta}}}{-im_{\hat{\pi}}\Gamma^{\mbox{\scriptsize{tot}}}_{
\hat{\pi}}}\right|^2 \Gamma_{\hat{\pi}\to3\pi}\approx0.023\
\mbox{MeV}\,. \end{eqnarray} Here we put $m_{\hat{\pi}}\approx1300$
MeV and $\Gamma^{\mbox{\scriptsize{tot}}}_{\hat{\pi}}\approx
\Gamma_{\hat{\pi}\to3\pi}\approx300$ MeV (note that the data on the
$\pi(1300)$ is very poor \cite{PDG2020}). This value is close to the
estimates of $\Gamma^{\mbox{\scriptsize{dir}} }_{\hat{\eta}\to\pi^0
\pi^+\pi^-}$ and $\Gamma^{\mbox{\scriptsize{dir}}}_{\hat{\eta}\to3
\pi^0}$ indicated in Eq. (\ref{Eq11}) for the mechanisms of the
direct $\eta(1295)\to3\pi$ decays. Let us try to determine the
widths of the individual decay modes that add up to $\Gamma^{
\mbox{\scriptsize{mix}}}_{\hat{\eta}\to \hat{\pi}^0\to3\pi}$. With
the help of isotopic invariance alone, which gives
\begin{eqnarray}\label{Eq13} \Gamma_{\hat{\pi}^+\to3\pi}=
\Gamma_{\hat{\pi}^+\to\pi^+\pi^+\pi^-}+\Gamma_{\hat{\pi}^+
\to\pi^+\pi^0\pi^0}=\Gamma_{\hat{\pi}^0\to3\pi}=
\Gamma_{\hat{\pi}^0\to\pi^0\pi^+\pi^-}+\Gamma_{\hat{\pi}^0
\to3\pi^0}\,,\nonumber \\ \Gamma_{\hat{\pi}^+\to\pi^+\pi^+\pi^-}=
\Gamma_{\hat{\pi}^+ \to\pi^+\pi^0\pi^0}+\Gamma_{\hat{\pi}^0
\to3\pi^0}\ \ \ (\mbox{or}\ \ \Gamma_{\hat{\pi}^0\to\pi^0\pi^+
\pi^-}=2\Gamma_{\hat{\pi}^+\to\pi^+\pi^0\pi^0})\,,
\end{eqnarray} this cannot be done.
A simple estimate of the components of $\Gamma^{
\mbox{\scriptsize{mix}}}_{\hat{\eta}\to\hat{\pi}^0\to3\pi}$ can be
obtained if we assume that the decay $\pi(1300)\to3\pi$ occurs via
$\rho\pi$ and $\sigma\pi$ intermediate states. In so doing, the
possible contributions of the $\pi\pi$ $S$-wave with isospin $I=2$
to the final three-pion states should be neglected. Experimental
data on the ratio between the $\rho\pi$ and $\sigma\pi$ modes are
inconsistent \cite{PDG2020,PdA17,Sa04,Ab01}. By analogy with Eq.
(\ref{Eq2}), we write the decay amplitude $\pi^0(1300)\to\pi^0\sigma
\to\pi^0(\pi^+\pi^- )_{{\scriptsize S}}$ in the form $T^{\hat{\pi}
}_{\sigma}(u)=C_{ \hat{\pi}}\,T^0_0(u)$. The naive quark model
allows us to relate the constants $C_{\hat{\pi}}$ and
$C_{\hat{\eta}}$ by the relation $C_{\hat{\pi}}=
C_{\hat{\eta}}/\sin( \theta_i-\theta_p)\approx1.376 C_{\hat{\eta}}$,
if, due to the proximity of the masses of $\pi(1300)$ and $\eta(1295
)$ states, we accept for the $\eta(1295)$ the quark structure of the
form $(u\bar u+d\bar d)/\sqrt{2}$. In that case the numerical
calculation similar to those done in Secs. II and III gives
$\Gamma_{\hat{\pi}^0\to\pi^0\sigma\to\pi^0\pi^+\pi^-}\approx85$ MeV
and $\Gamma_{\hat{\pi}^0\to\pi^0\sigma\to3\pi^0}\approx107$ MeV.
Consequently, approximately 200 MeV in the $\pi(1300)\to3\pi$ decay
width falls on the contribution from the $\sigma\pi$ intermediate
state and the rest is due to the $\pi(1300)\to\rho\pi\to3\pi$ decay.
For $\Gamma_{\hat{\pi}\to3\pi}\approx300$ MeV, the value of
$\Gamma^{\mbox{\scriptsize{mix}}}_{\hat{\eta}\to \hat{\pi}^0\to
3\pi}$ indicated in Eq. (\ref{Eq12}) is thus added from three
approximately equal partial widths
$\Gamma^{\mbox{\scriptsize{mix}}}_{\hat{\eta}\to \hat{\pi}^0\to
\pi^0\sigma\to\pi^0\pi^+\pi^-}$, $\Gamma^{\mbox{\scriptsize{mix}}
}_{ \hat{\eta}\to\hat{\pi}^0\to\pi^0\sigma\to3\pi^0}$, and
$\Gamma^{\mbox{\scriptsize{mix}}}_{ \hat{\eta}\to\hat{\pi}^0
\to\rho^\pm\pi^\mp\to\pi^0\pi^+\pi^-}$. Note that the decay mode
$\eta(1295)\to\rho^\pm\pi^\mp\to\pi^0\pi^+\pi^-$ appears only due to
the $\pi^0(1300)-\eta(1295)$ mixing.

\begin{figure} [!ht] 
\begin{center}\includegraphics[width=7cm]{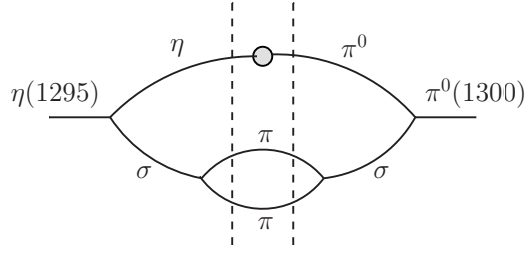}
\caption{\label{Fig6} An example of a diagram contributing to
Im$\Pi_{\hat{\pi}^0 \hat{\eta}}$. The vertical dashed lines cutting
the diagram mean that the 4-momenta of the intermediate particles,
either $3\pi$ or $\eta\pi\pi$, lie on their mass
shells.}\end{center}\end{figure}

The imaginary part of the transition amplitude $\Pi_{\hat{\pi}^0
\hat{\eta}}$ is due to the contributions of real intermediate
states. An example of a diagram that contributes to
Im$\Pi_{\hat{\pi}^0\hat{\eta}}$ is shown in Fig. \ref{Fig6}, where
the vertical dashed lines cutting the diagram mean that the
4-momenta of the intermediate particles, either $3\pi$ or
$\eta\pi\pi$, lie on their mass shells.
An estimate of this contribution to Im$\Pi_{\hat{\pi}^0\hat{\eta}}$
gives
\begin{eqnarray}\label{Eq14} \mbox{Im}\Pi_{\hat{\pi}^0\hat{\eta}}=
\frac{\Pi_{\pi^0\eta}}{m^2_\eta-m^2_{\pi^0}}\frac{C_{\hat{\eta}}C_{
\hat{\pi}}}{\pi M^2}\frac{3}{2}\left(\,\int\limits_{3\pi\,
\mbox{p.s.}}|T^0_0(u) |^2dsdu\ -\int\limits_{\eta\pi\pi\,
\mbox{p.s.}}|T^0_0(u)|^2dsdu \right)\nonumber\\
\approx(-0.0017+0.0004)\ \mbox{GeV}^2\approx -0.0013\
\mbox{GeV}^2.\qquad\qquad\qquad\end{eqnarray} Here, in the first
term, integration is carried out over the three-pion phase space
($3\pi\, \mbox{p.s.}$), and in the second, over the phase space of
$\eta\pi\pi$. The imaginary parts of the transition amplitudes
$\eta(1295)\to a^\pm_0(980)\pi^\mp\to\eta\pi^+\pi^-\to\pi^0
\pi^+\pi^-\to(\pi^0\sigma+\rho^\pm\pi^\mp)\to\pi^0(1300)$ and
$\eta(1295)\to a^0_0(980)\pi^0\to(K^+K^-+K^0\bar K^0)\pi^0\to
f_0(980)\pi^0\to\pi\pi\pi^0\to(\pi^0\sigma+\rho^\pm\pi^\mp)\to\pi^0(1300)$
cannot be easily estimated because of the need to take into account
contributions not only from three-body, but also from five-body real
intermediate states. However, they cannot greatly exceed the above
estimate of the contribution to Im$\Pi_{\hat{\pi}^0 \hat{\eta}}$
from the transition $\eta(1295)\to\eta \sigma\to\pi^0\sigma\to\pi^0
(1300)$. This is also confirmed by the upper estimate that roughly
takes into account all contributions to $|\mbox{Im}\Pi_{\hat{\pi}^0
\hat{\eta}}|$:
\begin{eqnarray}\label{Eq15} |\mbox{Im}\Pi_{\hat{\pi}^0\hat{\eta}}|<
m_{\hat{\eta}}\left(\sqrt{\Gamma^{\mbox{\scriptsize{dir}}}_{\hat{\eta}
\to\pi^0\pi^+\pi^-}\Gamma_{\hat{\pi}^0\to\pi^0\pi^+\pi^-} }+\sqrt{
\Gamma^{\mbox{\scriptsize{dir}}}_{\hat{\eta}\to3 \pi^0}\Gamma_{
\hat{\pi}^0\to3\pi^0}}\,\right)\nonumber \\ <
m_{\hat{\eta}}\left(\sqrt{\Gamma^{\mbox{\scriptsize{dir}}}_{\hat{\eta}
\to\pi^0\pi^+\pi^-}}+\sqrt{\Gamma^{\mbox{\scriptsize{dir}}}_{
\hat{\eta}\to3\pi^0}}\,\right)\sqrt{\Gamma_{\hat{\pi}^0\to3\pi}}
\approx0.0076\ \mbox{GeV}^2.\quad\end{eqnarray} So the estimates in
Eqs. (\ref{Eq11}), (\ref{Eq12}), (\ref{Eq14}), and (\ref{Eq15}) say
that the value of $\Gamma_{\eta(1295)\to 3\pi}$ may well turn out to
be of the order of $0.1$ MeV and, respectively, $\mathcal{B}(\eta
(1295)\to3\pi)\approx0.2\%$. In so doing, one can hope that in the
interference phenomena indications on the existence of the decays
$\eta(1295)\to\pi^0\pi^+\pi^-$ and $\eta(1295)\to3\pi^0$ can be
detected at a level of a few percent of the main $3\pi$ signal.

For $3\pi$ production cross sections ($\pi^0\pi^+\pi^-$ or $3\pi^0$)
in the pseudoscalar channel, we have
\begin{eqnarray}\label{Eq16} d\sigma(3\pi)=
\left|\frac{A_{\hat{\pi}^0}F_{\hat{\pi}^0\to3\pi}(s,t,u)}{D_{\hat{\pi}^0}(M)}
+\frac{A_{\hat{\eta}}[F^{\mbox{\scriptsize{dir}}}_{\hat{\eta}\to3\pi}(s,t,u)+
F^{\mbox{\scriptsize{mix}}}_{\hat{\eta}\to\hat{\pi}^0\to3\pi}(s,t,u)]}
{D_{\hat{\eta}}(M)}\right|^2dsdu\,,
\end{eqnarray}
where $A_{\hat{\pi}^0}$, $A_{\hat{\eta}}$ and $1/D_{\hat{\pi}^0}(M)
$, $1/D_{\hat{\eta}}(M)$ are the production amplitudes and the
Breit-Wigner propagators of the $\pi^0(1300)$, $\eta(1295)$
resonances, respectively, $F_{\hat{\pi}^0\to3\pi}(s,t,u)$ is the sum
of the amplitudes of the $\pi^0(1300)$ decay into $3\pi$ via
$\sigma\pi$ and $\rho\pi$ intermediate states, and
\begin{eqnarray}\label{Eq17} F^{\mbox{\scriptsize{mix}}}_{\hat{\eta}\to
\hat{\pi}^0\to3\pi}(s,t,u)=\frac{\Pi_{\hat{\pi}^0\hat{\eta}}
F_{\hat{\pi}^0\to3\pi}(s,t,u)}{D_{\hat{\pi}^0}(M)}\,.\end{eqnarray}
If the channel $3\pi^0$ is investigated, then the contribution of
the $\rho\pi$ mode is absent. If the channel $\rho^\pm\pi^\mp\to
\pi^0\pi^+\pi^-$ is separated, then  Eq. (\ref{Eq16}) does not
contain the contribution of the amplitude
$F^{\mbox{\scriptsize{dir}}}_{\hat{\eta}\to3\pi}(s,t,u)$. Since
$\Gamma^{\mbox{\scriptsize{tot}}}_{\hat{\eta}}$ is 4-6 times smaller
than $\Gamma^{\mbox{\scriptsize{tot}}}_{\hat{\pi}}$ \cite{PDG2020},
the signal from the $\eta(1295)$ resonance has some enhancement. Of
course, the interference pattern in the $\eta(1295)$ region depends
fundamentally on the relative magnitude of the amplitudes
$A_{\hat{\eta}}$ and $A_{\hat{\pi}^0}$.

\section{\boldmath{WHERE TO SEARCH FOR $\eta(1295)\to3\pi$ DECAYS?}}

The $J/\psi$ radiative decays are dominated by hadron production in
the  states with the isospin $I=0$. Therefore, the $\eta(1295)$
meson can manifest itself in the $J/\psi \to\gamma\eta(1295)\to
\gamma3\pi$ decays without accompaniment of the $\pi(1300)$. As
already noted in the Introduction, a hint at the $J/\psi\to \gamma
f_1(1285)/\eta(1295)\to\gamma\pi^0\pi^+\pi^-,\gamma \pi^0\pi^0\pi^0$
decays was obtained by the BESIII Collaboration \cite{Ab12} In this
experiment, the invariant masses of the $\pi^+\pi^-$ and $\pi^0
\pi^0$ pairs in the $\pi^0\pi^+\pi^-$ and $\pi^0\pi^0\pi^0$ mass
spectra were in the $f_0(980)$ region ($0.94$ GeV $<m_{\pi^+
\pi^-(\pi^0\pi^0)}<1.04$ GeV). In our model, the $\pi^+\pi^-$ and
$\pi^0 \pi^0$ masses spectra in this region are dominated by the
transition $\eta(1295)\to a^0_0 (980)\pi^0\to f_0(980)\pi^0 \to3\pi$
caused by $a^0_0(980)-f_0 (980)$ mixing, see Figs. \ref{Fig2}(b) and
\ref{Fig3}. Narrow peaks in the $\pi^+\pi^-$ and $\pi^0\pi^0$ mass
spectra are a good indicator of the $a^0_0(980)-f_0(980)$ mixing
mechanism (or in the general case of the $K\bar K$ loop
isospin-breaking mechanism \cite{AKS16,AS19}). One can hope that
searches for the signals from the $\eta(1295)$ resonance in the
decays $J/\psi\to\gamma\eta(1295)\to\gamma\eta\pi\pi$ and $J/
\psi\to\gamma \eta(1295)\to\gamma 3\pi$ will be successful.

Information about the $\eta(1295)\to3\pi$ decays can also be
obtained from interference experiments. For example, in the
semileptonic decay $D^+(c\bar d)\to d\bar de^+\nu_e\to3\pi e^+\nu_e$
the $d\bar d$ virtual intermediate state is not has a definite
isospin and can be a source of the $\eta(1295)$ and $\pi(1300)$
resonances with approximately equal production amplitudes
$A_{\hat{\eta}}$ and $A_{\hat{\pi}^0}$. Then the wavelike distortion
of the $\pi(1300)$ peak in the three-pion channel can make up
$\approx\pm5\%$ due to interference with the contribution from the
decay $\eta(1295)\to3\pi$. Of course, the observation of such
interference phenomena implies the availability of good data on the
main signal from the $\pi(1300)$ resonance. To obtain them, any
reactions can be involved, for example, those in which the
$\pi(1300)$ resonance was already observed earlier, i.e., peripheral
reactions, nucleon-antinucleon annihilation, $D$ meson decays,
$\gamma\gamma$ collisions, etc. \cite{PDG2020,AM20,PdA17,Ab01,
Sa04}. Methods of the partial-wave analysis are now well developed
and with sufficient statistics the extraction of $3\pi$ events
related to the pseudoscalar channel is although not a simple but
purely technical challenge.


\section{\boldmath{CONCLUSION}}

Studies of radial excitations of light pseudoscalar mesons are of
physical interest. The available data on the $\eta(1295)$, $\pi(13
00)$ and $K(1460)$ states are rather poor \cite{PDG2020}. In this
paper we attract attention of experimenters to the comprehensive
study of the $\eta(1295)$ state in its decay channels into $\eta
\pi\pi$ and $3\pi$. In the series of pseudoscalar isoscalar mesons
$\eta$, $\eta'$, $\eta(1295)$, and $\eta(1405)$ \cite{PDG2020} the
$\eta(1295)$ remained the last one that has not yet presented
unexpected surprises associated with the violation of isotopic
invariance in its decays into $\pi^0\pi^+\pi^-$ and $3\pi^0$. We
have constructed a simple isobar model for the description of the
main decay of the $\eta(1295)$ into $\eta\pi\pi$ and, based on this
model, estimated the widths of the direct decays
$\eta(1295)\to\pi^0\pi^+\pi^-$ and $\eta(1295)\to3\pi^0$ caused by
the mixing of the $\pi^0-\eta$ and $a^0_0(980)-f_0( 980)$ mesons.
Then we discussed a possible role of the $\pi^0(1300)-\eta(1295)$
mixing and obtained rough estimates for the $\eta(1295)\to3\pi$
decay widths with taking into account this additional mechanism of
the isospin breaking.
One can expect that the decay width $\eta(1295)\to3\pi$ will be of
the order of $0.1$ MeV and, respectively, $\mathcal{B}(\eta(1295)
\to3\pi)\approx 0.2\%$. One can hope that in the interference
phenomena indications on the existence of the decays
$\eta(1295)\to\pi^0\pi^+\pi^-$ and $\eta(1295)\to3\pi^0$ can be
detected at a level of a few percent of the main $3\pi$ signal. The
presented estimates are not overestimated. Finally, we have
discussed the processes that can be used for experimental searches
of the decays $\eta(1295)\to3\pi$. In particular, we have noted
reactions $J/\psi\to\gamma \eta (1295) \to\gamma 3\pi$ and
$D^+\to[\pi(1300)+\eta(1295)]e^+\nu_e \to 3\pi e^+\nu_e$.

Study of the decays $\eta(1295)\to\eta\pi\pi$, $\eta(1295)\to3\pi$,
and also $\pi(1300)\to3\pi$ can be a good challenge for new
high-statistics experiments. 

\begin{center} {\bf ACKNOWLEDGMENTS} \end{center}

The work was carried out within the framework of the state contract
of the Sobolev Institute of Mathematics, Project No. 0314-2019-0021.


\end{document}